\documentclass[11pt, times,twocolumn]{article}

\usepackage{graphicx}
\usepackage{subfig}
\usepackage{amssymb}

\begin{document}
\topmargin 0pt
\oddsidemargin 0mm
\renewcommand{\thefootnote}{\fnsymbol{footnote}}
\begin{titlepage}

\vspace{5mm}

\begin{center}
{\Large \bf Strain-tunable band gap in graphene/h-BN hetero-bilayer} \\

\vspace{6mm} {\large Harihar Behera\footnote{E-mail: harihar@phy.iitb.ac.in;  
 harihar@iopb.res.in } and Gautam Mukhopadhyay\footnote{Corresponding author's E-mail: gmukh@phy.iitb.ac.in; g.mukhopa@gmail.com}}    \\
\vspace{5mm}
{\em
Department of Physics, Indian Institute of Technology, Powai, Mumbai-400076, India} \\

\end{center}

\vspace{5mm}
\centerline{\bf {Abstract}}
\vspace{5mm}
  Using full-potential density functional calculations within local density 
approximation (LDA), we predict that mechanically tunable band-gap and 
quasi-particle-effective-mass are realizable in graphene/hexagonal-BN 
hetero-bilayer (C/h-BN HBL) by application of in-plane homogeneous biaxial 
strain. While providing one of the possible reasons for the experimentally 
observed gap-less pristine-graphene-like electronic properties of C/h-BN HBL, which theoretically has a narrow band-gap, we suggest a schematic experiment for 
 verification of our results which may find applications in 
nano-electromechanical systems (NEMS), nano opto-mechanical systems (NOMS) and 
other nano-devices based on C/h-BN HBL. \\


\end{titlepage}

\section{Introduction}
Recent experiments \cite{1,2} at Columbia University showing the successful 
fabrication of hexagonal BN (h-BN) gated graphene field effect transistors 
(BN-GFETs), consisting of both  mono-layer graphene (MLG) and bilayer graphene 
(BLG) on h-BN gate/substrate which could be made arbitrarily thin (down to a 
mono-layer of h-BN), are among the important developments in the physics and material science of graphene \cite{3,4,5,6}. Graphene supported on h-BN exhibits superior electrical 
properties with performance levels comparable to those observed in suspended 
samples \cite{1,2,6,7}.  On the other side, in the context of strain 
engineering of electronic properties of graphene, another significant 
experimental development \cite{8} utilizes piezoelectric actuators to apply 
tunable biaxial compressive as well as tensile stresses to graphene on h-BN 
substrate, which allows a detailed study of the interplay between the graphene 
geometrical structures and its electronic properties. In this scenario, it is 
useful and interesting to explore the effect of biaxial strain on the band gap, Fermi-velocity and the quasi-particle-effective-mass of a graphene/h-BN hetero-bilayer 
 (representing the smallest and the simplest form of BN-GFET), which is not only feasible but also desirable, since 
strain-engineering \cite{9} of Si, SiGe, Ge has been successfully used in the 
semiconductor industry to impressively improve the performance levels of 
conventional metal-oxide-semiconductor field-effect transistors (MOSFETs), 
and recently \cite{10} strain is seen as a solution for higher carrier mobility 
in nano MOSFETs. \\
\indent
   In this paper, we theoretically explore the effect of symmetry preserving 
homogeneous in-plane biaxial strain (which mimics an experimental 
  condition \cite{8} in which a flexible substrate supported graphene can be 
  strained in a controllable way by piezoelectric actuators) on a hetero-bilayer 
  (C/h-BN HBL) consisting of an MLG (C) and a mono-layer of h-BN, which is the 
  smallest and simplest form of BN-GFET. Previous theoretical studies on 
  graphene on h-BN substrate \cite{11} and C/h-BN HBL \cite{12,13} reported the
   opening of a small energy band gap in graphene which varies with (a) the 
   stacking order, 
   (b) the separation distance of graphene from the h-BN, and (c) the externally 
   applied perpendicular electric field. Other graphene/h-BN heterostructures 
   such as MLG \cite{14} and BLG \cite{15} sandwiched between two mono-layers of
    h-BN have also been theoretically shown to have (external) 
    electric-field-tunable electronic properties. However, a detailed study of 
    strain-engineered band gap of C/h-BN HBL is not available, although the 
    effects of various strain distributions on mono-layer h-BN \cite{16} and 
    graphene \cite{17,18,19,20} have been reported recently. Theoretically it is 
    shown that the high band gap of mono-layer h-BN is strain-tunable \cite{16},
     the gap-less nature of graphene is robust against small and moderate 
     deformations \cite{17,18,19,20}. On the experimental side, Raman 
     spectroscopy studies \cite{8,21,22,23,24} of graphene reveal that both 
     biaxial \cite{8} and uni-axial \cite{21,22,23,24} strains affect  
     the Raman Peaks; the transport properties of strained graphene have been 
   investigated by depositing samples on stretchable substrates \cite{25}. 
   Insights from these studies suggest a detectable strain-controlled band-gap 
   in a composite structure like C/h-BN HBL. This expectation has, in fact, been 
   borne out by the results of our present study. It is shown that the ground 
   state direct band gap ($\simeq 59$ meV) of C/h-BN HBL increases with biaxial 
   tensile strain and decreases with compressive strain, whereas the Fermi 
   velocity of charge carriers follows a reverse trend. Moreover, the variation 
   of computed average effective masses ($m^\star$) of charge carriers 
   (electrons or holes) with   strain is more pronounced with increasing 
   trend for tensile strain and decreasing trend for compressive strain. 
   These should affect the carrier mobility, conductivity and optical properties 
   of C/h-BN HBL which may be verified experimentally for which we propose 
   schematically a realistic experimental set-up.   
 
\section{Computational Methods}
Our first-principles study of electronic structure of C/h-BN HBL is based on 
full-potential (linearized) augmented plane wave plus local orbital 
(FP-(L)APW+lo) method\cite{26}, which is a descendant of FP-LAPW method\cite{27}. 
We use the elk-code\cite{28} and the Perdew-Zunger variant of LDA\cite{29}, 
the accuracy of which has been successfully tested in our previous work \cite{30}
 on graphene and silicene (silicon analog of graphene). The plane wave cut-off 
 of $|{\bf G}+{\bf k}|_{max} = 8.0/R_{mt}$ (a.u$.^{-1}$) ($R_{mt} =$ the 
 smallest muffin-tin radius) was chosen for plane wave expansion in the 
 interstitial region. The Monkhorst-Pack \cite{31} $k-$point grid of 
 $20\times20\times1$ is used for structural optimization and of 
 $30\times30\times1$ for electronic calculations of C/h-BN BLG with 
 simulated homogeneous in-plane biaxial strains up to $\pm 15\%$. Our 
 consideration of strains up to $15\%$ is motivated by recent theoretical 
calculations $[32-34]$ as well as experiments $[35]$ which demonstrated 
that graphene can sustain in-plane tensile elastic strain in excess of $20\%$; 
Kim et al. $[25]$ have measured resistances of graphene films transferred to pre-strained and unstrained PDMS substrates with respect to uni-axial tensile 
strain ranging from $0\%$ to $30\%$. Although in-plane homogeneous biaxial 
strains up to $\pm 15\%$ have been considered theoretically in $[36]$, the maximum compressive strain graphene can sustain is unclear to us. However, for the sake of symmetry and as a matter of curiosity, we have considered strains up to $-15\%$. The total energy was converged within $2\mu$eV/atom. We simulate the 2D-hexagonal 
 structure of C/h-BN BLG as a 3D-hexagonal super-cell with a large value of 
 $c$-parameter ($c = 40$ a.u.) by considering the primitive vectors of the unit cell as 
 \begin{equation}
 {\bf{a}} \,= (1/2)\,a\,{\bf{e_x}} - (\sqrt{3}/2)\,a\,{\bf{e_y}} 
 \end{equation}
 \begin{equation}
 {\bf{b}} \,= (1/2)\,a\,{\bf{e_x}} + (\sqrt{3}/2)\,a\,{\bf{e_y}} 
 \end{equation} 
 \begin{equation}
 {\bf{c}} \,= c\,{\bf{e_z}} 
 \end{equation}
 where $a$ is the in-plane lattice parameter and other symbols have their usual 
 meanings. The value of $a$ was varied to simulate the application of homogeneous 
 in-plane biaxial stress (since $|{\bf a}|$ = $|{\bf b}|$ = a) to C/h-BN HBL. Figure 1 schematically shows the structure of C/h-BN HBL in which (a) represents a typical structure, (b)-(d) represent top-down views of three different configurations.
\begin{figure}
 \subfloat[A typical C/h-BN HBL]{
 \label{fig:subfig:a} 
 \includegraphics[scale=0.75]{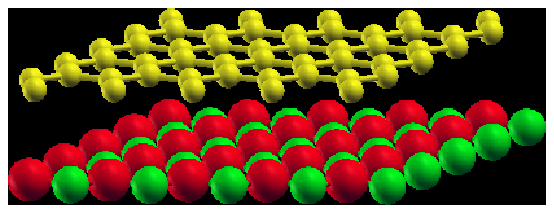}}
 \subfloat[B1]{
 \label{fig:subfig:b} 
 \includegraphics[scale=0.9]{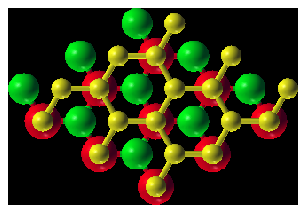}}
  \vspace{0.1in}
 \subfloat[B2]{
 \label{fig:subfig:c} 
\includegraphics[scale=0.9]{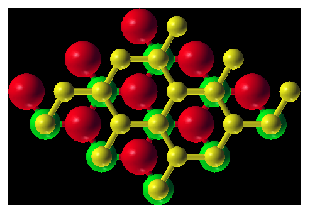}}
\hspace{0.6in}
 \subfloat[B3]{
 \label{fig:subfig:d} 
 \includegraphics[scale=0.9]{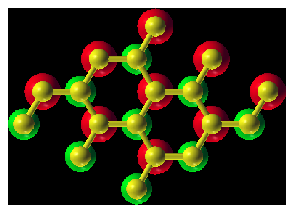}}
\caption{Ball-stick-model of C/h-BN HBL with top-down views of three different
configurations B1, B2 and B3. (a) A typical structure; (b) B1, one C atom (small ball
(yellow)) directly above the B atom (big ball (red)) and other C atom above the
center of h-BN hexagon; (c) B2, similar to B1 with B replaced by N (medium ball
(green)); (d) B3, one C atom is directly above the B and the other C atom directly
above N. (For interpretation of the references to color in this figure legend, the
reader is referred to the web version of this article.)
}
 \label{fig:subfig} 
\end{figure}
\begin{figure}
 \includegraphics[scale=0.8]{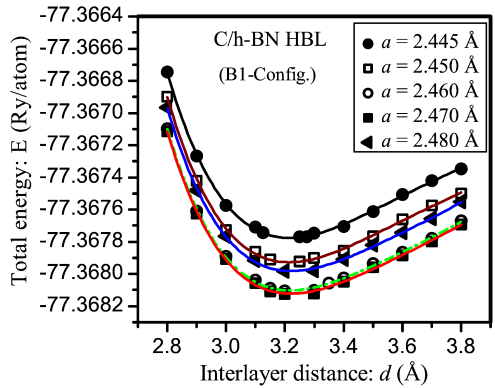}
 \caption{Total energy (E) vs interlayer distance $d$ at different $a$ values of C/h-BN HBL(B1).}
\end{figure} 
\section{Results and Discussions}
 Our calculated LDA lattice constants of graphene and h-BN mono-layer 
 respectively turned out as $a_0$(C) = 2.445 \AA \,and $a_0$(BN) = 2.488 \AA, 
 which are in excellent agreement with the reported values for graphene 
 ($a_0$(C) = 2.445 \AA \, in Ref. (11), 2.4426  \AA \, in Ref (17), 2.4431 \AA \, 
 in Ref (37)) and mono layer h-BN ($a_0$(BN) = 2.488 \AA \, in Ref. (16), 2.4870 \AA \, 
 in Ref (37)) based on plane wave pseudo-potential method. Graphene turned out
  as gap-less, while mono-layer h-BN with its valence-band maximum and the conduction 
  band minimum located at K point of the Brillouin Zone (BZ), revealed a direct 
  band gap of 4.606 eV, in agreement with previous calculations of 4.613 eV in 
  Ref. \cite{16}, 4.35 eV in Ref. \cite{37}. Our estimated band gap energy of h-BN mono-layer
   is about 23\% less than the experimental \cite{38} direct band gap energy of 
   5.971 eV.\\
  
\begin{figure}
 \includegraphics[scale=0.8]{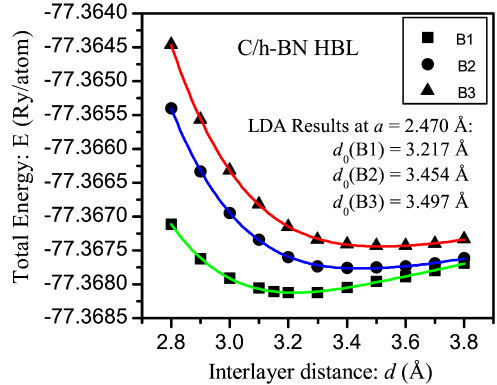}
 \caption{Variation of E with $d$ at a common value of 
 $a = 2.47$ \AA \, for B1, B2 and B3 configurations of C/h-BN HBL.}
\end{figure} 
 \begin{figure}
 \includegraphics[scale=0.8]{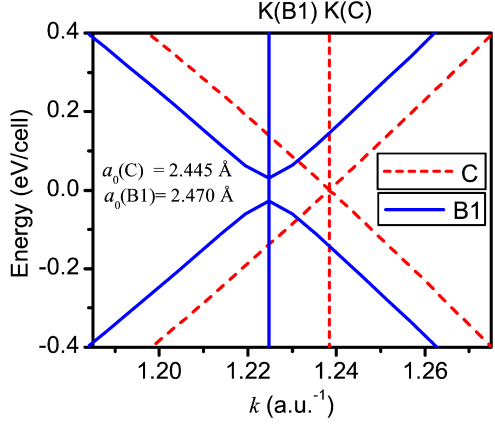}
 \caption{Comparison of energy bands (within LDA) of graphene (C) and 
 C/h-BN HBL(B1) near the K point of the BZ. Fermi energy $\mbox{E}_F = 0.0$ eV.}
\end{figure} 
\begin{figure}
 \includegraphics[scale=1.0]{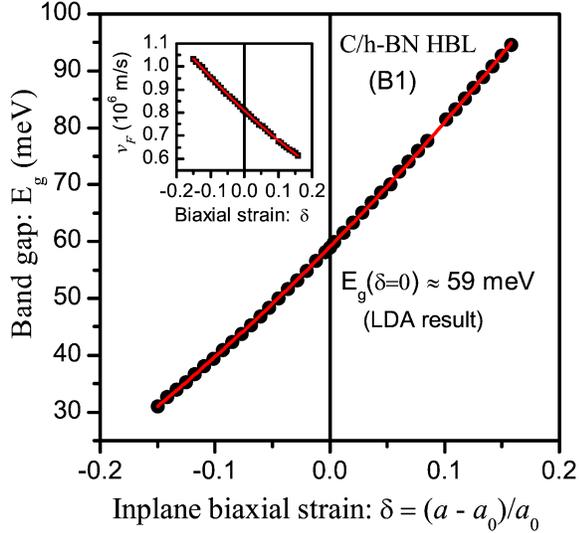}
 \caption{Variation of E$_{\mbox{g}}$ of C/h-BN HBL(B1) with $\delta $.
  Inset shows the variation of Fermi velocity $v_F$ with $\delta$.}
\end{figure}
\begin{figure}
 \includegraphics[scale=1.0]{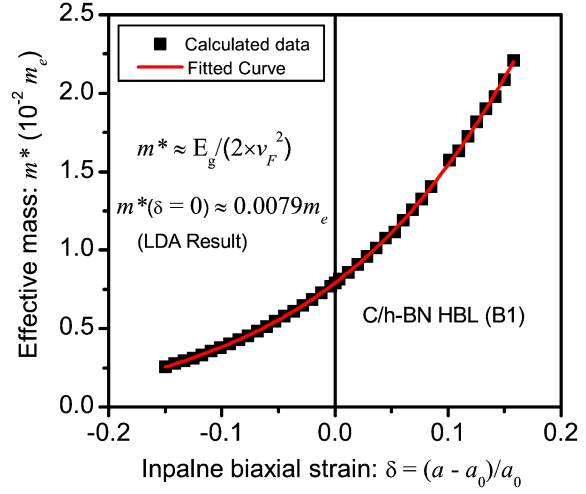}
 \caption{Variation of effective mass ($m^\star$) of charge carriers in C/h-BN HBL(B1) with $\delta$.}
\end{figure}

 \indent
 The calculations for the composite structure of C/h-BN HBL were done as 
  follows, unlike the earlier authors \cite{11,12} who used the LDA value of 
  lattice constant of graphene ($a_0$(C)= 2.445 \AA) for the 
  three configurations of C/h-BN HBL. 
  As per the previous studies \cite{11,12}, B1 configuration of C/h-BN HBL 
  (henceforth we call it C/h-BN HBL(B1) or simply B1) is the most stable 
  configuration with equilibrium interlayer spacing $d_0$(B1) = 3.22 \AA \, at 
  an assumed $a_0$(B1) = 2.445 \AA. For B1, with which we are concerned in 
the present paper, the calculated ground state total 
  energies at different “$d$” values for five fixed values of “$a$” ( = 2.445, 2.450,
   2.460, 2.470, 2.480 \AA \,) were obtained as depicted in Figure 2; from this 
  we estimated the optimized values of “$a$” and “$d$” as  $a_0$(B1) = 2.47 \AA \,  
  and $d_0$(B1) = 3.217 \AA. Similar calculations for B2 and B3 are required for 
  a comparison of the relative stability of B1, B2 and B3. However, like 
  previous authors \cite{11,12,13}, we got the relative stability and minimum 
  interlayer distance of B1, B2 and B3 at constant “$a = 2.47$” \AA \, 
  as depicted in Figure 3, which corroborate the reported results\cite{11,12,13}. 
  The interlayer spacing of B2 and B3 at a common value of  $a$ = 2.47 \AA \, 
  were estimated as $d_0$(B2) = 3.454 \AA, $d_0$(B3) = 3.497 \AA \, which agree 
  with the reported results \cite{11,12} of  $d_0$(B2) = 3.4 \AA \, and 
  $d_0$(B3) = 3.5 \AA. \\
  \indent  
  The band structures of unstrained graphene and unstrained C/h-BN HBL(B1)
  near the K point of the BZ are depicted in Figure 4 for comparisons. 
  As seen in Figure 4, graphene is gap-less, whereas C/h-BN HBL(B1) has a 
  direct band gap of E$_{\mbox{g}}$= 59 meV which is of same order of 
  magnitude  as that of graphene on a hexagonal BN 
  substrate \cite{11} (E$_{\mbox{g}}$ = 53 meV) in the same configuration. 
  Therefore, the number of h-BN layers below graphene mono-layer has little 
  impact on the band structure of C/h-BN heterostructure. 
  It is to be noted that E$_{\mbox{g}}$ is under-estimated here because of the LDA. 
  However, in experiments $[2]$ involving graphene on h-BN substrates, there is 
no evidence of this gap from transport measurements and this absence of gap has 
been attributed to randomly stacked graphene on h-BN $[2]$. The fabrication of a 
graphene/h-BN heterostructure with a desired alignment seems to be a technological 
hurdle at present. However, we note that Fan et al. $[13]$ have investigated one 
type of misalignment of graphene to h-BN layer (along with the three aligned structures 
B1, B2 and B3 we have considered), which they have achieved by translating BN 
mono-layer a distance of ${\sqrt 3}a/6$ ($a$ is the lattice constant of graphene) 
with respect to graphene along the C-C bond orientation from a pattern that corresponds 
to the B3 configuration of our present study. This misaligned heterostructure with 
an estimated equilibrium interlayer distance of $3.4$ \AA\, is energetically 
shown $[13]$ to be unstable with respect to the most stable configuration that 
corresponds to the B1 configuration of our present study. Further, 
this misaligned structure is pictorially shown to have a small band gap which 
vanishes when the interlayer distance reaches $4.0$ \AA. It is to be noted that the 
theoretical study in $[13]$ is concerned with the effect of interlayer 
spacings on electronic structure of graphene/h-BN hetero-bilayer and 
there is no consideration of biaxial strain effects. We also note that 
since our simulated homogeneous biaxial strain preserves the hexagonal 
symmetry of an aligned C/h-BN HBL (say B1), a misalignment is not expected to occur 
in such C/h-BN HBL under homogeneous biaxial strain which only affect the size but not the shape of the hexagonal structure under study.\\
\indent
 The energy dispersion around K point of BZ is linear for graphene and 
  therefore the low energy quasi-particles mimic the massless Dirac fermion 
  behavior\cite{3,4}, i.e.,  
\begin{equation}
\mbox{E}_{\pm}\,\simeq \mbox{E}_F\,\pm \hbar v_F k 
\end{equation} 
where $\mbox{E}_F$ is Fermi energy, $v_F$ is Fermi velocity of quasi-particles, 
and $(\hbar k)$ is the momentum. In case of B1, low energy dispersion is not only linear
 in $k$ but also parallel to the low energy dispersion of graphene, while very 
 low energy dispersion within an energy window of (-0.06 eV , +0.06 eV) around 
 $\mbox{E}_F$ (= 0.0 eV) seems quadratic. Since the major portion of the energy 
 dispersions around the K point (Dirac point) have identical slopes 
 (hence identical $v_F$ values) for graphene and C/h-BN HBL(B1), 
 we attribute this to one of the reasons for B1 (in spite of having a small energy gap) behaving like free standing graphene as observed experimentally. 
  From the linear energy dispersion portion of the plot, the calculated average 
  value of $v_F$ turned out as $0.8\times10^6$ m/s for both graphene and 
  C/h-BN HBL(B1), in agreement with other calculated 
  result\cite{13} and in close proximity of the experimental value $\approx 10^6$ m/s 
  for graphene \cite{3,4,39}. Assuming that the 
   effective mass ($m^\star$) of charge carriers of C/h-BN HBL at Dirac point 
   given by the relation \cite{13} $m^\star \simeq E_g/2{v_F}^2$, we estimated 
   $m^\star \simeq 0.0079\mbox{m}_{\mbox{e}}$ for B1 
   (where $\mbox{m}_{\mbox{e}}$ is the free electron mass), which is much smaller than 
   the effective Dirac fermions’ mass of 0.03$\mbox{m}_{\mbox{e}}$ in bilayer 
   graphene \cite{40}. The high value of $v_F$ (equal to the $v_F$ value 
   in graphene), a very low value of $m^\star$ can render C/h-BN HBL pristine 
   graphene-like electronic properties. However, it is to be noted that the observed pristine graphene-like electronic properties of C/h-BN HBL(B1) can not be attributed to these factors alone. This is due to the fact that experimentally the graphene/h-BN systems are fabricated by mechanical ex-foliation and transfer, so the stacking between graphene and BN can not be pre-determined and relaxed in such techniques. 
  \\
 \indent 
For C/h-BN HBL(B1), the calculated variation of (i) E$_{\mbox{g}}$
  with in-plane homogeneous biaxial strain $\delta = (a-a_0)/a_0$ is depicted in 
  Figure 5, the inset shows the variation of $v_F$ with $\delta$;  (ii) $m^\star$ 
  with $\delta$ is depicted in Figure 6.\\   
 \indent
 Our predicted strain-induced modifications of E$_{\mbox{g}}$, $v_F$ and $m^\star$
  should affect the transport and optical properties of C/h-BN HBL, which may be probed 
  experimentally in a set-up (thought of as a specific amalgamation of the reported experimental \cite{1,2,8} designs) like the one schematically shown in Figure 7. 
  Although further improvements in the design and materials of Figure 7 are 
  possible, the important message is that such an experiment is feasible with 
  the available technology as applied successfully in the experiments 
  \cite{1,2,8}. In Figure 7, a simple h-BN back-gated BN-GFET\cite{1,2} 
  is rigidly fixed on a variable biaxial strain inducing system (VSIS) like 
  the one described in Ref. \cite{8}. By applying a definite voltage across the VSIS, 
  a biaxial strain of definite magnitude can be induced in BN-GFET. The 
  current-voltage characteristics of the back-gated BN-GFET at different strain 
  conditions may be studied to see to what extent the biaxial strain affects 
  the transport properties of BN-GFET. Moreover, an optical probe of the direct
   band-gap of C/h-BN HBL may be performed at different strained states of 
   BN-GFET in such a set-up. This opto-mechanical probe may ascertain if C/h-BN 
   HBL can be used in designing nano opto-mechanical systems (NOMS).
 \begin{figure}
 \includegraphics[scale=0.9]{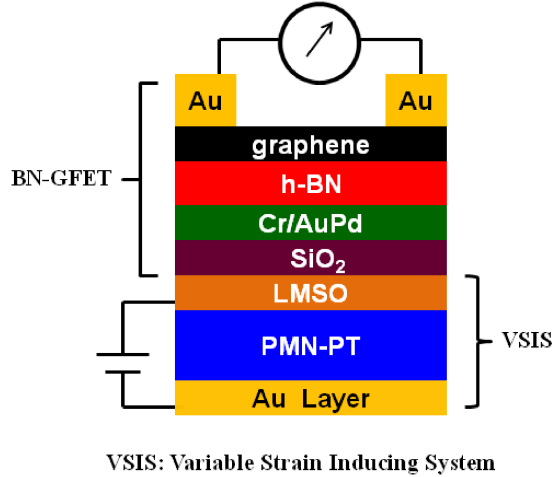}
 \caption{Schematic view of the proposed experimental set-up to study the 
 strain-induced modifications of the electronic properties of C/h-BN HBL. 
 Specific materials and fabrication schemes may be found in the experiments \cite{1,8}.}
\end{figure}
 \section{Conclusions} 
 In conclusion, we have studied and compared the band structures of graphene and 
 C/h-BN HBL (in biaxially strained and unstrained states) using density 
 functional theory based FP-(L)APW+lo method. From the parallel slopes of the 
 energy dispersion bands near K point of the BZ of these structures, we 
 estimated identical values of Fermi velocity of the low energy 
 charge carriers in them. From this and an estimated small value of 
 effective mass $m^\star \simeq 0.0079\mbox{m}_{\mbox{e}}$, 
 we infer the pristine graphene-like electronic properties of 
 narrow-gapped C/h-BN HBL. However, these results do not necessarily 
 reflect the measured transport properties of actual devices. In particular, 
 the other important factors determining the actual device performance 
 include the flatness of the substrate, surface phonon energy and 
 scattering, and the trap charges as in the oxide substrates. The 
 strain-induced modifications of the energy band gap, Fermi velocity and 
 effective mass of charge carriers in C/h-BN HBL in its most stable 
 configuration was studied in detail with output of some testable results 
 for which a schematic experimental set-up has been proposed. 
 The results, if verified, may find applications in future 
 graphene-based NEMS, NOMS and other nano-devices. 


\begin{thebibliography}{00}
\bibitem{1} I. Meric, C. Dean, A. Young, J. Hone, P. Kim, K.L. Shepard, 
Technical Digest International Electron Devices Meeting, IEDM, 2010, 23.2.1.  
\bibitem{2} C.R. Dean, A.F. Young, I. Meric, C. Lee, L. Wang, S. Sorgenfrei, 
K. Watanabe, T. Taniguchi, P. Kim, K.L. Shepard, J. Hone, Nat. Nanotechnol. 5 (2010), 722.
\bibitem{3} A.K. Geim, Science 324 (2009) 1530.
\bibitem{4} A.H. Castro Neto, F. Guinea,N.M.R. Peres, K.S. Novoselov, A.K. Geim, Rev.
     Mod. Phys. 81 (2009) 109. 
\bibitem{5} N.M.R. Peres, Rev. Mod. Phys. 82 (2010) 2673.
\bibitem{6} S.D. Sarma, E.H. Hwang, Phys. Rev. B 83 (2011)121405(R). 
\bibitem{7} R.T. Weitz, A. Yacoby, Nat. Nanotechnol. 5 (2010) 699. 
\bibitem{8} F. Ding, H. Ji, Y. Chen, A. Herklotz, K. D\"{o}rr, Y. Mei, 
A. Rastelli, O.G. Schmidt, Nano Lett. 10 (2010) 3453. 
\bibitem{9} M. L. Lee, E.A. Fitzgerald, M.T. Bulsara, M.T. Currie, A. Lochtefeld,
 J. Appl. Phys. 97 (2005) 011101.
\bibitem{10} M. Chu, Y. Sun, U. Aghoram, S.E. Thompson, Annu. Rev. Mater. Res. 
39 (2009) 293.  
\bibitem{11} G. Giovannetti, P.A. Khomyakov, G. Brocks, P.J. Kelly, van den J. Brink, 
Phys. Rev. B 76 (2007) 073103. 
\bibitem{12} J. S\l awi\'{n}ska, I. Zasada, Z. Klusek, Phys. Rev. B. 81 (2010) 155433.
\bibitem{13} Y. Fan, M. Zhao, Z. Wang, X. Zhang, H. Zhang, Appl. Phys. Lett. 98 (2011) 083103.
\bibitem{14} J. S\l awi\'{n}ska, I. Zasada, P. Kosi\'{n}ski, Z. Klusek, Phys. 
Rev. B 82 (2010) 085431. 
\bibitem{15} A. Ramasubramaniam, D. Naveh, E. Towe, Nano Lett. 11 (2011) 1070.
\bibitem{16} J. Li, G. Gui, J. Zhong, J.  Appl. Phys. 104 (2008) 094311.   
\bibitem{17} R.M. Ribeiro, V.M. Pereira, N.M.R. Peres, P.R. Briddon, 
A.H. Castro Neto, New J. Phys. 11 (2010) 115002.
\bibitem{18} M. Farjam, H. Rafii-Tabar, Phys. Rev. B 80 (2009) 167401.
\bibitem{19} G. Gui, J. Li, J. Zhong, Phys. Rev. B 80 (2009) 167402.
\bibitem{20} Seon-M. Choi, Seung-H. Jhi, Young-W. Son, Phys. Rev. B 81 (2010) 081407(R).
\bibitem{21} Z.H. Ni, T. Yu, Y.H. Lu, Y.Y. Wang, Y.P. Feng, Z.X. Shen, ACS Nano 2 (2008) 2301.
\bibitem{22} T.M.G. Mohiuddin, A. Lombardo, R.R. Nair, A. Bonetti, G. Savini, R. Jalil, 
N. Bonini, D.M. Basko, C. Galiotis, N. Marzari, K.S. Novoselov, A.K. Geim, 
A.C. Ferrari, Phys. Rev. B 79 (2009) 205433.
\bibitem{23} M. Huang, H. Yan, T.F. Heinz, J. Hone, Nano Lett. 10 (2010) 4074.
\bibitem{24} O. Frank, M. Mohr, J. Maultzsch, C. Thomsen, I. Riaz, R. Jalil, 
K.S. Novoselov, G. Tsoukleri, J. Parthenios, K. Papagelis, L. Kavan, C. Galiotis, 
ACS Nano 5 (2011) 2231. 
\bibitem{25} K.S. Kim, Y. Zhao, H. Jang, S.Y Lee, J.M. Kim, K.S. Kim, Jong-H. Ahn, 
P. Kim, Jae-Y. Choi, B.H. Hong, Nature 457 (2009) 706.  
\bibitem{26} E. Sj\"{o}stedt, L. Nordstr\"{o}m, D.J. Singh, Solid State Commun. 
114 (2000) 15. 
\bibitem{27} E. Wimmer, H. Krakauer, M. Weinert, J.A. Freeman, Phys. Rev. B 24 (1981) 864.
\bibitem{28} Elk is an open source code: http://elk.sourceforge.net/  
\bibitem{29} J.P. Perdew, A. Zunger, Phys. Rev. B 23 (1981) 5048. 
\bibitem{30} H. Behera, G. Mukhopadhyay, AIP Conf. Proc. 1313 (2010) 152.
\bibitem{31} H.J. Monkhorst, J.D. Pack, Phys. Rev. B 13 (1976) 5188. 
\bibitem{32} F. Liu, P. Ming, J. Li, Phys. Rev. B 76 (2007) 064120.
\bibitem{33} E. Cadelano, P. L. Palla, S. Giordano, L. Colombo, Phys. Rev. Lett. 102 
(2009) 235502.
\bibitem{34} X. Wei, B. Fragneaud, C. A. Marianetti, J. W. Kysar, Phys. Rev. B 80 (2009) 205407. 
\bibitem{35} C. Lee, X. Wei, J. W. Kysar, J. Hone, Science 321(5887) (2008) 385.
\bibitem{36} G. Cocco, E. Cadelano, L. Colombo, Phys. Rev. B 81 (2010) 241412.
\bibitem{37} S. Wang, J. Phys. Soc. Jpn. 79 (2010) 064602.   
\bibitem{38} K. Watanabe, T. Taniguchi, H. Kanda, Nat. Mater. 3 (2004) 404.
\bibitem{39} K.S. Novoselov, A.K. Geim, S.V. Morozov, D. Jiang, M.I. Katsnelson,
 I.V. Grigorieva, S.V. Dubonos, A.A. Firsov, Nature 438 (2005) 197.
\bibitem{40} E.V. Castro, N.M.R. Peres, J.M.B.L. dos Santos, F. Guinea, 
A.H. Castro Neto, J. Phys.: Conf. Ser. 129 (2008) 012002. 
\end{thebibliography}
\bibliographystyle{elsarticle-num}

\end{document}